\documentclass[MSNbibl,nameyear,dvips]{arxstspdf}
\usepackage{flushend}
\usepackage{stfloats}

\volume{26}
\issue{2}
\pubyear{2011}
\firstpage{257}
\lastpage{259}
\doi{10.1214/11-STS346A}
\referstodoi{10.1214/10-STS346}

\begin{document}
\begin{frontmatter}

\title{Discussion of
``Impact of Frequentist and Bayesian Methods on Survey Sampling
Practice:
A Selective Appraisal'' by J.~N.~K.~Rao}
\runtitle{Discussion}
\pdftitle{Discussion of Impact of Frequentist and Bayesian Methods on
Survey Sampling Practice: A Selective Appraisal by J. N. K. Rao}

\begin{aug}
\author[a]{\fnms{Glen} \snm{Meeden}\corref{}\ead[label=e1]{glen@stat.umn.edu}}
\runauthor{G. Meeden}

\affiliation{University of Minnesota}

\address[a]{Glen Meeden is Professor,
School of Statistics,
University of Minnesota,
224 Church St. SE,
Minneapolis, Minnesota 55455-0493, USA \printead{e1}.}

\end{aug}



\end{frontmatter}

It has been pointed out that when apologists for competing systems
like capitalism and socialism or  the frequentist and Bayesian approaches to
survey sampling argue about
the relative merits of their systems, they often compare their ideal to
the other's reality. Since  the ideal  is always quite different than reality
it is easy for each of them to score points.
I~wish to thank  Rao for avoiding this trap
and giving a fair reading to both sides in his survey. Beyond that,
I particularly liked the sections on the early development of frequentist
methods.

How should prior information about the population be used in survey sampling?
It can inform how the sample is selected and is used when making  inferences
after the data have been collected. Formally, at each of the two stages,
 the frequentist and Bayesian approaches
  are quite different but practically, I believe, they are often
more alike than is commonly supposed.

\section*{The Frequentist Approach}

In theory, for non-model-based frequentists, the sampling design is the
most important place to
 use   prior information. In Section 2  Rao described some of the
early fundamental advances in  survey design based on the frequentist approach.
 He explained why in stratified
sampling and in stratified two-stage cluster sampling,  where one
cluster within each stratum is drawn, self-weighting of the units is a very desirable
property. Such examples led to the notion of assigning a
weight, which is the reciprocal of its inclusion probability, to each unit in the
sample. A~unit's weight is the number of units in the population that
it represents. A  theoretical justification for this notion that is often given
is that under the sampling design the resulting estimator
is unbiased. Rao argues, however, that large sample consistency
of an estimator is a more important property than unbiasedness.
Although it is hard to find sensible estimators which
are badly biased, I agree with him that unbiasedness in and of itself is not an important
property. Whatever justification there is for the notion of a weight, it should not
be based on unbiasedness.

What I have sometimes  found puzzling  about\break weights is that after the sample has been
selected they are often adjusted.  Information that may
not have been used at the design stage is used to make the sampled units and
 their weights   more accurately reflect
what is known about the population of interest.
Calibration and the model-assisted approach are two common methods
for achieving this end.  An  estimator based on the adjusted weights will no
longer be design-unbiased,  but there is theory to show that it can be  design-consistent.  Practice, however, can be more complicated especially when there
are missing observations. But more importantly, the whole reweighting technology
 seems to me to mix up an unconditional argument
(selecting  the sampling  design)
 and a conditional argument (using population information to get a good estimate after
the sample has been observed). I am not suggesting that such adjustments should
not be done, only that there can be more art than science in finding  a good set of weights.

I believe that frequentists would be better served   in their analysis if they more
explicitly recognized these two different stages in the inferential process.
In the  first stage, one uses  the information that is relevant  to select
the sampling design. In the second stage, after the sample has been selected,
 one should ignore the design but
use \textit{all} the information when constructing an estimator. In
effect that is what one does when the sampling  weights are adjusted.
However, in the second stage, the
design weights need not be used explictly as long as
all the information is being taken into account. If all the information
has been used wisely, then the resulting estimate should work well
whenever this particular sample has been observed no matter how it was selected.
Again,  I~am not suggesting that
much of current frequentist practice  is badly flawed.  But  I believe
  design-based practitioners  should realize
that when trying to decide how to select a good sample they are  arguing
unconditionally and when adjusting the weights  at the estimation stage they
are arguing conditionally  and in the latter stage they should pay less explicit
attention to the design.

\section*{The Bayesian Approach}

The usual Bayesian approach requires the specification of a prior distribution over the
possible population values.
Basu argued (Basu, \citeyear{ghosh88}),
 correctly I believe, that for a Bayesian after the sample has been observed,
how it was selected should not enter into the analysis. This assumes of course that
any information used in selecting the design was also available to the Bayesian when
choosing the prior distribution. This does not mean that
a Bayesian is indifferent to how a sample is selected. In theory a~Bayesian should use the prior distribution to select an optimal, purposeful sample. However, as  a practical
matter this almost never happens. The reason is that the typical kinds of prior
information available in the finite population
setting seldom  lend themselves to summarization in a prior distribution.
 The most common situation where a Bayesian
can find an optimal sample is when the prior distribution is exchangeable
and any sample is just as good or bad as any other. Therefore it seems unlikely to me
that the Bayesian approach will be useful when deciding how to select a  sample.
Meeden and Noorbaloochi (\citeyear{m-n10})  argue, however, that in some situations the sampling design can be
thought of as part of the prior distribution.

 After the sample is observed, inferences for a Baye\-sian are in theory straightforward.
Given a sample, one uses the posterior distribution to
 simulate many\vadjust{\eject} complete copies of the
population. For each simulated complete copy one calculates the population
quantity of interest, say a mean or a quantile.
 The average of these   simulated values is their point estimate
and the 0.025 quantile and the 0.975 quantile of the values forms an approximate 0.95
credible interval. As Rao noted, Bayes methods have proven useful in
small area estimation where certain model assumptions lead to
fairly simple hierarchal priors.
 The more general lack  of the utilization of Bayesian methods, despite these attractive
features,  results from two factors.  Specifying sensible prior distributions
can be very difficult, and even with the recent advances in Markov chain Monte Carlo
methods  simulating complete copies of the
population from a posterior distribution can also  be difficult.

Consider situations where,  given a sample,
the statistician believes the unsampled or unseen units are
like the sampled or seen units.  This
 happens, for example,  under simple random sampling.
In such cases the Polya posterior (PP) yields a nonparametric objective
pseudo/Bayesian justification for many of the standard methods.
The PP  does not arise from a single prior distribution but is actually a~family
of posteriors that arise from a sequence of
prior distributions. Ghosh and Meeden (\citeyear{ghomee97})  give the underlying
stepwise Bayes theory for this approach which proves the admissibility of
many of the standard estimators.

The stepwise Bayes theory allows for a more flexible Bayesian-like approach.
Rather than selecting a~single prior before the sample is chosen,
one selects a~posterior, after the sample has been observed, which
uses the sample and all available information about the population to relate
the unseen to the seen.
 One needs to verify that all the posteriors fit together in a stepwise Bayes
manner to guarantee the admissibility of the resulting procedure for any design.
Even so, this can be much easier  to do than specifying a~single prior.

Lazar, Meeden and Nelson (\citeyear{l-m-n08}) showed how population information about auxiliary variables can
be incorporated
into  the PP after  a sample has been observed. Examples include knowing the
population mean or median of  auxiliary  variables  and more generally  only
knowing  that they fall in some known intervals. No model assumptions are made
about how the auxiliary variables are related to the variable of interest.
For this constrained version of the PP, the R (RDC Team, \citeyear{r05}) package \textit{polyapost}
is available to generate simulated copies of the complete population.

\section*{Final Remarks}

 I believe that once a sample has been selected
the key issue is how the unseen are related to the seen.
I believe   that this is in line with much  frequentist practice
although  this is  obscured by the prominent and unnecessary role played
by the design weights after the sample has been selected.
For a~Bayesian, with a prior distribution, this happens automatically through
the posterior distribution, but has been of limited value in practice.
 I believe that the stepwise Bayesian approach should make it easier
to select useful posteriors which make use of all the prior information present.
But as Rao pointed out, this approach  needs to be extended to more complicated
sampling designs.

In most of sample survey, given a design, any procedure,
be it frequentist or Bayesian, should be evaluated by how it behaves under
repeated sampling from  the  design. For point estimators either their
average mean squared error loss or average absolute error loss is the
quantity of interest.  Rather than focusing on getting an estimate of
variance for the  estimator to measure its precision,
one should focus on using the estimator to find approximate 95\% confidence intervals
for the parameter of interest.\

\vspace*{-3pt}

\end{document}